\documentstyle[prl,aps]{revtex}

\batchmode

\begin{document}
\author{Yu. Kagan, L.A. Maksimov}
\address{RRC ''Kurchatov Institute'', Kurchatov Sq. 1, 123182 Moscow, Russia}
\title{Manifestation of superfluidity in an evolving Bose-condensed gas}
\date{Tue Aug 01 2000}
\maketitle

\begin{abstract}
We study the generation of excitations due to an ''impurity''(static
perturbation) placed into an oscillating Bose-condensed gas in the
time-dependent trapping field. It is shown that there are two regions for
the position of the local perturbation. In the first region the condensate
flows around the ''impurity'' without generation of excitations
demonstrating superfluid properties. In the second region the creation of
excitations occurs, at least within a limited time interval, revealing
destruction of superfluidity. The phenomenon can be studied by measuring the
damping of condensate oscillations at different positions of the
''impurity''.
\end{abstract}

\ In studies of Bose-Einstein condensation in ultracold gases there has been
outlined a new stage associated with investigating superfluidity (SF) in
such systems. This problem is of a specific interest since it is a question
of SF of a dilute gas. The first results in revealing a critical velocity
have been obtained in \cite{1}. At the same time in \cite{2} the creation of
a vortex is demonstrated in a sophisticated experiment with two condensates.
Recently, direct creation of vortices has been observed in a rotating
condensate \cite{3}.

An interesting possibility for studying SF may be realized in the
observation of evolution of condensate accompanying time-dependent
variations of the confining field. For the most interesting case when a
parabolic potential keeps its shape, the scaling solutions found in \cite
{4,5,6} allow us to describe the evolution of condensate for arbitrary
variations of the trap frequencies. The analysis of solutions, in
particular, given below, shows that during the evolution there are
space-time regions where the velocity $v(\vec{r},t)$ of condensate exceeds
the local speed of sound $c(\vec{r},t)$, i.e., the local critical velocity
for the SF system. Simultaneously, there exists a spatial region where, on
the contrary, $v(\vec{r},t)<c(\vec{r},t)$ for any time. In the absence of
external perturbations the presence of the regions with $v(\vec{r},t)>c(\vec{%
r},t)$ does not result in itself in the irreversible processes and the
oscillations of condensate do not decay at a temperature $T=0.$

However, provided an external perturbation is localized in the region where $%
v(\vec{r},t)>c(\vec{r},t)$ even for a limited time interval, generation of
collective excitations of the condensate takes place. Naturally, this
results in the decay of the oscillations of the condensate. At the same
time, if the perturbation is located in the region where always $v(\vec{r}%
,t)<c(\vec{r},t)$ , the gas in the presence of SF will flow adiabatically
about the local perturbation not creating excitations and, therefore,
without decay of the oscillations of the condensate.\ Thus, measuring the
decay of oscillations of the condensate under changing localization of
perturbation, we can display the SF properties of a Bose-condensed gas.

As is found in \cite{4}, for the isotropic 2D parabolic potential there is
an exact scaling solution for arbitrary values of the gas density $n$ and
initial trap frequency $\omega _{0}$ and an arbitrary dependence $\omega
\left( t\right) $. Moreover, the energy of the interparticle interaction and
kinetic energy in this case change identically with an expanding gas,
entailing similar variation of the main parameters. On the whole, this makes
a choice of the trap geometry close to the cylindric one very attractive. In
this case a static perturbation can be induced by the laser beam parallel to
the longitudinal cylindric axis and limited in the transverse size. Shifting
the beam axis in the radial direction, one can change the localization of
perturbation.

It is worthwhile to note that for this type of investigations, a trapped
Bose- condensed gas has an essential advantage in comparison with the $%
^{4}He $ quantum liquid. This is related with the isolation of a gas from
any walls, the possibility to concentrate a majority of particles in the
condensate, and with studying the appearance of irreversibility and the
partial destruction of coherence at the conditions of the large-scale
time-spatial evolution of condensate with the strong change of density

Below we consider the evolution of the condensate in an isotropic 2D
parabolic potential with the time-dependent frequency $\omega \left(
t\right) $ and determine the generation of excitations under external static
perturbation. The generalization for the 3D case of the cylindric symmetry
and extended perturbation unchanging along the longitudinal axis can be
performed easily.

Examining evolution of the 2D interacting Bose gas in the time-dependent
confining potential, we can employ the general equation for the Heisenberg
field operator of atoms

\begin{equation}
i\hbar \frac{\partial \hat{\Psi}(\vec{r},t)}{\partial t}=[-\frac{\hbar ^{2}}{%
2m}\Delta +\frac{m\omega ^{2}(t)r^{2}}{2}]\hat{\Psi}(\vec{r},t)+U_{0}\hat{%
\Psi}^{+}(\vec{r},t)\hat{\Psi}(\vec{r},t)\hat{\Psi}(\vec{r},t).  \label{1a}
\end{equation}
The only simplification in the equation is an assumption of the local
character of the interparticle interaction which allows us to present this
interaction in the form $U\left( r_{1}-r_{2}\right) =U_{0}\delta \left(
r_{1}-r_{2}\right) $.

Let us introduce a scaling parameter $b\left( t\right) $ and,
correspondingly,\ spatial and time variables $\vec{\rho}=
\vec{r} b\left(t\right) ,\hspace{-1cm}\tau (t).$

For the 2D case, let us represent the operator $\hat{\Psi}\left( r,t\right) $
as 
\begin{equation}
\hat{\Psi}\left( r,t\right) =\frac{1}{b}\hat{\chi}\left( \vec{\rho},\tau
\right) \exp \left[ i\Phi \left( r,t\right) \right] .  \label{2a}
\end{equation}
Substituting (\ref{2a}) into (\ref{1a}) and using the results obtained in 
\cite{1}, we find 
\begin{equation}
\Phi \left( r,t\right) =mr^{2}\frac{1}{2\hbar b}\frac{db}{dt}.  \label{3a}
\end{equation}

Then the equation for operator $\hat{\chi}\left( \vec{\rho},\tau \right) $\
reduces to 
\begin{equation}
i\hbar \frac{\partial \hat{\chi}}{\partial \tau }=-\frac{\hbar ^{2}}{2m}%
\Delta _{\rho }\hat{\chi}+\frac{1}{2}m\omega _{0}^{2}\rho ^{2}\hat{\chi}%
+U_{0}\hat{\chi}^{+}\hat{\chi}\hat{\chi}.  \label{4a}
\end{equation}
The equation takes such form if we accept that $b\left( t\right) $\ and $%
\tau (t)$\ are determined by the equations 
\begin{equation}
\frac{d^{2}b}{dt^{2}}+\omega ^{2}\left( t\right) b=\frac{\omega _{0}^{2}}{%
b^{3}},\quad \tau (t)=\int\limits_{0}^{t}dt^{\prime }/b^{2}(t^{\prime }).
\label{5a}
\end{equation}
We put $\omega \left( -\infty \right) =\omega _{0}$ in Eqs. (\ref{4a}) and (%
\ref{5a}). Equation (\ref{5a}) should be solved at the initial conditions of 
$b\left( -\infty \right) =1,$ $\dot{b}\left( -\infty \right) =0$\ .

It follows from Eq.(\ref{4a}) that the problem in the $\vec{\rho}$ and $\tau 
$ variables reduces to the solution of the equation for the static parabolic
potential of the initial frequency $\omega _{0}.$ Finding solution and
deriving $b\left( t\right) $\ and $\tau \left( t\right) $\ from (\ref{5a}),
we obtain a complete description for the space-time evolution at an
arbitrary\ variation of $\omega \left( t\right) $. It is essential that this
refers, to the same extent, to both the condensate and the excited states of
the system.

Let us restrict our consideration with the case of zero temperature. Then
the ground state in the static potential represents the condensate and in
Eq.(\ref{4a}) the operator $\hat{\chi}$\ can be replaced by the macroscopic
condensate wavefunction $\chi _{0}\left( \vec{\rho},\tau \right) $\ having a
typical dependence on $\tau $%
\begin{equation}
\chi _{0}\left( \vec{\rho},\tau \right) =\chi _{0}\left( \vec{\rho}\right)
e^{-i\mu \tau /\hbar }.  \label{6a}
\end{equation}

Here $\mu $ is the starting chemical potential and $\chi _{0}\left( \vec{\rho%
}\right) $ is the solution of the equation 
\begin{equation}
-\frac{\hbar ^{2}}{2m}\Delta _{\rho }\chi _{0}+[-\mu +\frac{1}{2}m\omega
_{0}^{2}\rho ^{2}+U_{0}\chi _{0}^{2}]\chi _{0}=0.  \label{7}
\end{equation}

To describe excitations in the system, we introduce operator $\hat{\chi}%
^{\prime }\left( \vec{\rho},\tau \right) $. Let us write the initial
operator $\hat{\chi}$\ as $\hat{\chi}\left( \vec{\rho},\tau \right) \ =\
[\chi _{0}\left( \vec{\rho}\right) \ +\ \hat{\chi}^{\prime }\left( \vec{\rho}%
,\tau \right) ]\ e^{-i\mu \tau /\hbar }$ and substitute it into (\ref{4a}).
Then we find for the equation linearized in $\hat{\chi}^{\prime }\left( \vec{%
\rho},\tau \right) $%
\begin{equation}
i\hbar \frac{\partial \hat{\chi}^{\prime }}{\partial \tau }=-\frac{\hbar ^{2}%
}{2m}\Delta _{\rho }\hat{\chi}^{\prime }+(\frac{1}{2}m\omega _{0}^{2}\rho
^{2}-\mu )\hat{\chi}^{\prime }+2U_{0}\chi _{0}^{2}\hat{\chi}^{\prime
}+U_{0}\chi _{0}^{2}\hat{\chi}^{\prime +}.  \label{8a}
\end{equation}
This equation determines excitations in the coordinates of $\left( \vec{\rho}%
,\tau \right) $ and we can consider any problem in this reference frame.
However, keeping in mind the solution of the problem on creating excitations
under a static perturbation, it is more natural to investigate it in the
laboratory reference frame.

Let us rewrite Eq.(\ref{8a}) introducing new variables $\vec{r}$ and $t$.
For this purpose, we use the relations 
\[
\frac{\partial \hat{\chi}^{\prime }}{\partial \tau }=b^{2}\frac{\partial 
\hat{\chi}^{\prime }}{\partial t}+\vec{r}b\dot{b}\frac{\partial \hat{\chi}%
^{\prime }}{\partial \vec{r}},\quad \Delta _{\rho }\hat{\chi}^{\prime
}=b^{2}\Delta \hat{\chi}^{\prime }. 
\]

We assume that the density of a gas is sufficiently large and $\mu \gg \hbar
\omega _{0},$ i.e., the Thomas-Fermi approximation is valid. Then we find
the condensate density from Eq.(\ref{7}) (see (\ref{2a})) 
\begin{equation}
n_{0}\left( \vec{r},t\right) =\frac{1}{b^{2}\left( t\right) }\chi
_{0}^{2},\quad \chi _{0}^{2}=\frac{\mu }{U_{0}}(1-\frac{r^{2}}{R^{2}\left(
t\right) }),  \label{9a}
\end{equation}
where 
\begin{equation}
R\left( t\right) =R_{0}b\left( t\right) ,\quad R_{0}=\sqrt{2\mu /m\omega
_{0}^{2}}.  \label{10a}
\end{equation}

As a result, in the laboratory reference frame Eq.(\ref{8a}) reads 
\begin{equation}
i\hbar \frac{\partial \hat{\chi}^{\prime }}{\partial t}=-\frac{\hbar ^{2}}{2m%
}\Delta \hat{\chi}^{\prime }+g\hat{\chi}^{\prime }+g\hat{\chi}^{\prime
+}-i\hbar \vec{v}\nabla \hat{\chi}^{\prime }.  \label{11a}
\end{equation}

Here 
\begin{eqnarray}
\vec{v} &=&\vec{r}\frac{\dot{b}}{b},  \label{12a} \\
g &=&U_{0}n_{0}\left( \vec{r},t\right) =\frac{\mu }{b^{2}\left( t\right) }(1-%
\frac{r^{2}}{R^{2}\left( t\right) })  \label{13a}
\end{eqnarray}
It is easy to see that the quantity $\vec{v}$ is a local velocity of the
condensate. In fact, the total condensate wavefunction equals 
\begin{equation}
\Psi _{0}\left( r,t\right) =\frac{1}{b}\chi _{0}\exp \left[ i\Phi \left(
r,t\right) -\frac{i\mu \tau (t)}{\hbar }\right] .  \label{14a}
\end{equation}
Taking into account (\ref{3a}) and determining phase gradient, we arrive at
expression (\ref{12a})\ for the velocity. A special attention should be paid
for the appearance of the Doppler $\vec{v}\vec{p}\hat{\chi}^{\prime }$ term
in Eq.(\ref{11a}), $\hat{p}=\left( -i\hbar \nabla \right) $ being the
operator of momentum.

Note that, according to (\ref{9a}), the density$\ n_{0}\left( \vec{r}%
,t\right) $ is proportional to $b^{-2}\left( t\right) $. Correspondingly,
the correlation length of $\xi \sim n_{0}^{-1/2}$ changes in time as $\xi
=\xi _{0}b\left( t\right) $ where $\xi _{0}=\hbar \left( 2mU_{0}n_{0}\left(
0,0\right) \right) ^{-1/2}.$ The size of the system $R\left( t\right) $ (see
(\ref{10a})) changes in the same manner. Thus a ratio of these quantities $%
\xi \left( t\right) /R\left( t\right) =\xi _{0}/R_{0}$ conserves for an
arbitrary scale of evolution.

The inequality $\xi _{0}/R_{0}\ll 1$ is crucial for the Thomas-Fermi
approximation. So, if this approximation is valid at the initial time, it
remains valid for an arbitrary time. Since $\xi \left( t\right) /R\left(
t\right) \ll 1$ , we have a quasi uniform problem at any time, considering
excitations of the $\xi \left( t\right) \leq \lambda \ll R\left( t\right) $
wavelengths. The scale of the uniform regions, according to Eq.(\ref{11a}),
changes with the typical time of $t_{eff}\approx b\left( t\right) /\dot{b}%
\left( t\right) .$ The well-defined excitations satisfy the inequality of $%
\omega t_{eff}\gg 1$, i.e., satisfy the quasistationary condition.

Let us consider the case of the fast transition from frequency $\omega _{0}$
to $\omega _{1}=\omega _{0}/\beta ,\beta >1.$ In this case the solution of
Eq.(\ref{5a}) yields 
\begin{equation}
b^{2}\left( t\right) =\frac{1}{2}\left( \beta ^{2}+1\right) -\frac{1}{2}%
\left( \beta ^{2}-1\right) \cos 2\omega _{1}t.  \label{15a}
\end{equation}

Hence 
\begin{equation}
\vec{v}=\vec{r}\frac{\omega _{1}}{2b^{2}}\left( \beta ^{2}-1\right) \sin
2\omega _{1}t.  \label{16b}
\end{equation}
As we will see below, the excitations of $\lambda \sim \xi $ give the main
contribution into the damping. The energy of these excitations is of the
order of $U_{0}n_{0}\left( 0,t\right) =\mu /b^{2}\left( t\right) $ and,
correspondingly, 
\[
\omega t_{eff}\approx \frac{2\mu }{\hbar \omega _{0}}\frac{\beta }{\left(
\beta ^{2}-1\right) \sin 2\omega _{1}t}. 
\]
It is clear that the $\omega t_{eff}\gg 1$ condition holds if $2\mu /\hbar
\omega _{0}\gg \beta .$ The last inequality can easily be fulfilled. Thus,
we can seek for the solution of Eq.(\ref{11a}) within the quasiuniform and
quasistationary approximation.

Let us introduce a typical representation for operator $\hat{\chi}^{\prime }$%
\begin{equation}
\hat{\chi}^{\prime }/b=\sum\limits_{\vec{k}}\exp \left( i\vec{k}\vec{r}%
\right) \hat{a}_{\vec{k}},  \label{16a}
\end{equation}
where $\hat{a}_{\vec{k}}$ is the annihilation operator of a particle. We
employ the known Bogoliubov transformation, e.g.,\cite{7}, for the operator $%
\hat{a}_{\vec{k}}$%
\begin{equation}
\hat{a}_{\vec{k}}=u_{\vec{k}}\hat{b}_{\vec{k}}-v_{\vec{k}}\hat{b}_{-\vec{k}%
}^{+},  \label{17a}
\end{equation}
where $\hat{b}_{\vec{k}}^{+}$ and $\hat{b}_{\vec{k}}$ are the creation and
annihilation operators of collective Bose excitations. As a result, the
solution of Eq. (\ref{11a}) gives the excitation spectrum 
\begin{equation}
\varepsilon _{\vec{k}}=\hbar \vec{k}\vec{v}+\tilde{\varepsilon}_{\vec{k}%
},\quad \tilde{\varepsilon}_{\vec{k}}=\sqrt{E_{\vec{k}}^{2}-g^{2}},\quad E_{%
\vec{k}}=\frac{\hbar ^{2}\vec{k}^{2}}{2m}+g,  \label{18a}
\end{equation}
The expression for the coefficients of the transformation is given by 
\begin{equation}
u_{\vec{k}}=\sqrt{\frac{1}{2}\left( \frac{E_{\vec{k}}}{\tilde{\varepsilon}_{%
\vec{k}}}+1\right) },v_{\vec{k}}=\sqrt{\frac{1}{2}\left( \frac{E_{\vec{k}}}{%
\tilde{\varepsilon}_{\vec{k}}}-1\right) }.  \label{19a}
\end{equation}
Let an external perturbation of a size of about $d<<R_{0}$ and localized
near $\vec{r}=\vec{r}_{p}$ be described by an effective interaction $B(\vec{r%
}-\vec{r}_{p})$. The Hamiltonian of the interaction of the gas with the
perturbation has the form 
\[
H_{int}=\int d^{2}r\hat{\Psi}^{+}(\vec{r},t)B(\vec{r}-\vec{r}_{p})\hat{\Psi}(%
\vec{r},t)=\frac{1}{b^{2}}\int d^{2}r\hat{\chi}^{+}(\vec{r},t)B(\vec{r}-\vec{%
r}_{p})\hat{\chi}(\vec{r},t). 
\]
Selecting terms corresponding to the single-particle excitations, we have 
\begin{equation}
H_{int}=\frac{1}{b^{2}}\int d^{2}r[\chi _{0}(\vec{r},t)B(\vec{r}-\vec{r}_{p})%
\hat{\chi}^{\prime }(\vec{r},t)+hc\,].
\end{equation}
Let us transform this Hamiltonian, using (\ref{16a}), (\ref{17a}), and value
of $\chi _{0}=b\sqrt{n_{0}}$

\begin{equation}
H_{int}=\sqrt{n_{0}(\vec{r}_{p},t)}\sum_{\vec{k}}B_{\vec{k}}\exp \left( i%
\vec{k}\vec{r}_{p}\right) [u_{\vec{k}}-v_{\vec{k}}]\left( \hat{b}_{\vec{k}}+%
\hat{b}_{-\vec{k}}^{+}\right) .  \label{21a}
\end{equation}

If we take into account Eq.(\ref{18a}), the probability per unit time of
creating excitations equals ($T=0$) 
\begin{equation}
W=\frac{2\pi }{\hbar }n_{0}\left( \vec{r}_{p},t\right) \int \frac{d^{2}k}{%
\left( 2\pi \right) ^{2}}\left| B_{k}\right| ^{2}[u_{k}-v_{k}]^{2}\delta
\left( \tilde{\varepsilon}_{k}+\hbar \vec{k}\vec{v}\right) .  \label{22a}
\end{equation}

In accordance with Eq.(\ref{19a}) 
\[
(u_{k}-v_{k})^{2}=\frac{\hbar ^{2}k^{2}}{2m\tilde{\varepsilon}_{k}}. 
\]
Integration over the angles in Eq.(\ref{22a}) yields 
\[
\int d\varphi \delta \left( \hbar kv\cos \varphi +\tilde{\varepsilon}%
_{k}\right) =\frac{2\theta \left( 1-\left( \tilde{\varepsilon}_{k}/\hbar
kv\right) \right) }{\hbar kv\sqrt{1-\left( \tilde{\varepsilon}_{k}/\hbar
kv\right) ^{2}}}, 
\]
where $\theta \left( x\right) $ is the step function: $\theta \left(
x\right) =1$ for $x\geq 0$ and $\theta \left( x\right) =0$ for $x<0.$ Then 
\begin{equation}
W=\frac{n_{0}\left( \vec{r}_{p},t\right) }{2\pi vm}\int dk\left|
B_{k}\right| ^{2}\frac{k^{2}}{\tilde{\varepsilon}_{k}}\frac{\theta \left(
1-\left( \tilde{\varepsilon}_{k}/\hbar kv\right) \right) }{\sqrt{1-\left( 
\tilde{\varepsilon}_{k}/\hbar kv\right) ^{2}}}.  \label{23a}
\end{equation}
As follows from this expression, creation of an excitation with the
wavevector $k$ occurs provided $v>\tilde{\varepsilon}_{k}/\hbar k,$ i.e.,
provided that the familiar Landau criterium holds. The generation of
excitations is completely absent if $v\left( \vec{r}_{p},t\right) <c\left( 
\vec{r}_{p},t\right) $.

The local velocity of sound equals (see (\ref{18a})) 
\[
c\left( \vec{r},t\right) =\sqrt{\frac{g\left( \vec{r},t\right) }{m}}=\frac{%
c_{0}}{b\left( t\right) }\left( 1-\frac{r^{2}}{R^{2}\left( t\right) }\right)
^{1/2}, 
\]
where $c_{0}=\sqrt{\mu /m}.$ Taking into account the local magnitude of the
gas velocity (\ref{16b}) and expression (\ref{15a}), we have 
\begin{equation}
\alpha =\frac{v^{2}\left( \vec{r}_{p},t\right) }{c^{2}\left( \vec{r}%
_{p},t\right) }=\frac{2\eta ^{2}\left( b^{2}-1\right) \left( \beta
^{2}-b^{2}\right) }{\beta ^{2}\left( b^{2}-\eta ^{2}\right) }\theta \left(
b-\eta \right) ,\quad \eta =\frac{r_{p}}{R_{0}},\quad \beta =\frac{\omega
_{0}}{\omega _{1}}.  \label{25a}
\end{equation}
For $1<\eta <\beta $, the dependence of $\alpha $ on $b^{2}\left( t\right) $
is a monotonically decreasing function with the inevitably existing region
where the generation of excitations ($\alpha >1$) takes place. If $1<\eta
\ll \beta ,$%
\[
b_{\min }=\eta ,b_{\max }=\beta \left( 1-\frac{1}{4\eta ^{2}}\right) . 
\]

If $\eta <1,$ the value $\alpha $ as a function of $b^{2}$ has a maximum at
the point 
\[
b_{m}^{2}=\eta ^{2}+\sqrt{\left( \beta ^{2}-\eta ^{2}\right) \left( 1-\eta
^{2}\right) }. 
\]
Calculating $\alpha _{m}$ at this point, we can verify that the generation
of excitations does not appear in the whole region of $r_{p}<R_{0}$ if $%
\beta <\sqrt{2}.$ At the same time, the generation is absent at $r_{p}<R_{0}/%
\sqrt{2}$ for an arbitrary value of $\beta $.

Suppose that $r_{p}>R_{0}$ and let us calculate an integral in Eq.(\ref{23a}%
). The values of $k$ are limited by the condition which follows from the
equality\ $\tilde{\varepsilon}_{k}=\hbar kv$, 
\begin{equation}
k\leq k_{*},\quad k_{*}=\frac{2m}{\hbar }\sqrt{v^{2}-c^{2}}.  \label{27a}
\end{equation}

In the 2D case the effective vertex $B_{k}$ in the interaction Hamiltonian (%
\ref{21a}) at $k_{*}d<1$ can be represented as 
\begin{equation}
B_{k}=\frac{B_{k}^{\left( 0\right) }}{1+\left( mB_{k}^{\left( 0\right) }/\pi
\hbar ^{2}\right) \ln \left( 1/kd\right) },  \label{26a}
\end{equation}
where $B_{k}^{\left( 0\right) }$ is the Fourier transform of the bare
potential, $d$ being a length of the order of the size of perturbation
potential in the 2D case. If this value is large enough, we have 
\begin{equation}
B_{k}=\frac{\pi \hbar ^{2}}{m\ln \left( 1/kd\right) }  \label{31a}
\end{equation}
i.e., it has only a weak logarithmic dependence on $k$. In the opposite case 
$k_{*}d>1$ Eqs. (\ref{25a}) and (\ref{31a}) are not valid and the
calculation of the transition amplitude requires a knowledge of the concrete
form of the interaction potential. The most contribution into the integral (%
\ref{23a}) is given by the interval adjoined the upper limit $k_{*}$. We can
approximately take a factor of $\left| B_{k}\right| ^{2}$ out of the
integral, putting $k_{*}$ into the argument of the logarithm. Next, integral
(\ref{23a}) can be calculated exactly as an integral over $x=\left( \tilde{%
\varepsilon}_{k}/\hbar vk\right) $ within the limits $\left( c/v,1\right) $%
\begin{equation}
W=\frac{mn_{0}\left( \vec{r}_{p},t\right) \left| B_{*}\right| ^{2}}{\hbar
^{3}}[1-\frac{2}{\pi }\arcsin \frac{c}{v}]\theta \left( 1-\frac{c}{v}\right)
,\quad B_{*}=B_{k_{*}}.  \label{29a}
\end{equation}
Let us determine the total number of excitations created for a period of the
condensate oscillations in the case of $1<\eta \ll \beta .$

\begin{equation}
I=2\int\limits_{t_{1}}^{t_{2}}dtW.
\end{equation}

The time $t_{1}$ determines the moment when the boundary of the gas front
reaches the point $r_{p}$. This value is found from the condition $%
r_{p}=b\left( t_{1}\right) R_{0}.$ It follows from Eq.(\ref{15a}) that $%
b^{2}\left( t_{1}\right) \approx 1+\left( \omega _{0}t_{1}\right) ^{2}$ and $%
t_{1}\approx \sqrt{\eta ^{2}-1}/\omega _{0}\ll 1/\omega _{1}.$ The time $%
t_{2}$, found from the condition $\alpha =1$ (\ref{25a}), equals $%
t_{2}\approx [\pi /2-1/\left( \sqrt{2}\eta \right) ]/\omega _{1}$.
Substituting Eq.(\ref{9a}) for the density into (\ref{29a}), one can see
that the integral over time is dominated by the lower limit. From the
physical point of view this is associated with the fact that the condensate
density at point $r_{p}$ has a maximum during evolution just for $t\sim
t_{1} $. It follows from (\ref{25a}) that $c/v\leq 1/\left( \sqrt{2}\eta
\right) $. Under this condition the factor in the square brackets in Eq.(\ref
{29a}) is of the order of unity. After integration we find

\begin{equation}
I\approx \frac{1}{\hbar ^{3}}mn_{0}(0,0)\left| B_{*}\right| ^{2}\frac{1}{%
\eta \omega _{0}}.  \label{I1}
\end{equation}
Substituting (\ref{31a}) into (\ref{I1}) and introducing the total number of
particles $N=\frac{1}{2}\pi R_{0}^{2}n_{0}(0,0)$ instead of $n_{0}(0,0)$, we
find 
\begin{equation}
I\approx \frac{\pi \hbar \omega _{0}}{\eta \mu }\frac{N}{\ln ^{2}\left(
1/k_{*}d\right) },\quad \eta =\frac{r_{p}}{R_{0}}.  \label{31b}
\end{equation}

In the present case we have put $\hbar \omega _{0}/\mu \ll 1.$ This means
that the number of excitations $I$ and also the number of particles escaped
from the condensate for the oscillation period are small compared with $N.$

The total energy of the excitations generated during one oscillation period
can be estimated as $\mu \eta ^{2}I.$ Since the energy of the oscillating
condensate is about $\mu N$, the relative energy loss for a period at $\eta
\geq 1$ and $\beta \gg 1$ approximately equals $\eta ^{2}I/N$. Varying the
initial condensate density ($\mu $) or the 'impurity' position, one can
change the damping rate.

Treating a realistic problem in the quasi 2D case corresponding to the
cylindric symmetry of the field configuration with the transverse parabolic
potential and perturbation independent of the longitudinal coordinates, we
arrive at the result of (\ref{31b}) with $N$ equal to the total 3D number of
particles. One can readily\ see that, if the condensate 3D wavefunction is
written as $\Psi _{0}^{(3)}(x,y,z)=\Psi _{0}^{(2)}(x,y)/\sqrt{L_{z}}$ and
normalized by the total number of particles, the wavefunction $\Psi
_{0}^{(2)}$ is governed by the same Gross-Pitaevsky equation as in the
purely 2D case with the replacement of $U_{0}$ $\rightarrow
U_{0}^{(2)}\rightarrow U_{0}^{(3)}/L_{z}$. Since $%
n_{0}^{(2)}=L_{z}n_{0}^{(3)},$ we have $%
n_{0}^{(2)}U_{0}^{(2)}=n_{0}^{(3)}U_{0}^{(3)}.$ The cylindric symmetry of
perturbation results in creating collective excitations of $k_{z}=0$ alone.
In addition, the problem of scattering proves to be purely two-dimensional
and we arrive at Eqs.(\ref{31a}) and (\ref{31b}). The quantities $%
B_{k}^{(0)} $ and $B_{k}$ remain as 2D Fourier components. Note, that, if
the laser beam parallel to the cylindric axis is used as a static
perturbation (possibly, with the light sheets at the edges), an influence of
the regions where the beam enters and leaves the condensate is unessential.

In conclusion, we have investigated the generation of excitations in the
oscillating condensate in a time-dependent parabolic trap at the presence of
a static local perturbation. We have revealed the existence of two space
regions for the position of an ''impurity''. In the first region of $%
r_{p}<R_{0}/\sqrt{2}$ the generation of excitations is absent at any scale
of oscillations. In the second one of $r_{p}>R_{0}$ the generation is
realized inevitably for an arbitrary set of parameters and we have found the
generation rate. The results for the first region are a direct consequence
of SF of condensate. For any position in the second region, there is a
finite time interval when the local velocity of the condensate proves to be
larger than the local velocity of sound, entailing violation of SF.
Measurements of the damping of the condensate oscillations at different
positions of the perturbation opens a possibility to study the manifestation
of SF in the evolving condensate.

This work was supported by the Russian Foundation for Basic Research and by
the Grants INTAS-97-0972 and INTAS-97-11066. This work was partially
performed while one of the authors (Yu. K.) was at Technical University of
Munich under the Humboldt Research Award program.

\end{document}